# Coupling Thermal Integration of a Solid Oxide Fuel Cell with a Magnesium Metal Hydride Tank

E.I Gkanas[1,2], S.S Makridis[1,2], A.K Stubos[1]

[1] Institute of Nuclear and Radiological Sciences, Energy, Technology and Safety, NCSR "Demokritos", Aghia Paraskevi, Athens, 15310, Greece

[2] Institute for Renewable Energy and Environmental Technologies, University of Bolton, Deane Road, Bolton, BL3 5AB, UK

Abstract

The dehydriding behavior of a cylindrical Mg metal hydride tank is simulated and examined in the case where the tank is thermally coupled with an operating Solid Oxide Fuel Cell (SOFC) at $700^0$C. A three-dimensional validated mathematical model is utilized to simulate the hydrogen desorption from a cylindrical Mg hydride tank. Four scenarios are simulated: a base case where the heat source for the desorption process is an external heater surrounding the tank. The second case examines the effect of the radiation heat transfer from the SOFC to the metal hydride as a possible heat source for the desorption procedure. The third scenario uses the exhaust air from the SOFC cathode as the heating source which is driven to the hydride and the fourth scenario is a combination of both the exhaust air from the SOFC cathode and the external heater as the heat source for the desorption. According to the results, the exhaust air from the SOFC and the combination of external heater and the exhaust heat have a uniform temperature distribution within the tank and enhance the desorption capacity.

1. Introduction

Hydrogen storage in reversible metal hydrides is desirable for use in applications such as hybrid electric vehicles [1] because hydrogen can be stored at low pressures with a high volumetric density [2]. Further, hydrogen can be used essentially pollution free and produced from renewable energy resources, thus eliminating the net production of greenhouse gases. Hydrogen storage in metal hydrides is also particularly advantageous for stationary or small-scale fuel cell applications where the desire for small storage tank outweighs the disadvantages of the hydride mass [3].

The main drawback for the metal hydride storage systems, is the limited heat transfer between the external heat source (external water bath, spiral coil surrounding the tank, etc.), and the metal hydride within the tank, where the reaction takes place [4]. The heat transfer rate can determine the rate of hydrogen desorption [5] and by increasing these rates inside the hydride tanks is crucial for optimizing the design of hydrogen storage applications. Methods to increase the heat transfer by manipulating the internal properties of the storage vessels have been proposed. These methods include: Insertion of nickel or aluminum foam inside the tank [6, 7], integration of copper wire net structure [8], compacting metal hydride powders with expanded graphite [9] and internal heat exchangers [10].

When hydrogen is released from the metal hydride bed, the endothermic desorption process causes the temperature inside the metal hydride to decrease [11]. As a result, this temperature decrease inside the metal hydride tank diminishes the rate at which hydrogen releases from the tank. In order to eliminate the effect, some of the heat naturally produced from a Solid Oxide Fuel Cell (SOFC) during its operation can be transferred to the hydride bed in order to increase the temperature of the tank. In this way, the tank can thermally coupled with the SOFC, by using the exhaust waste heat from the SOFC and drive it to the tank or by using the irradiation heat transfer from the SOFC to the surface of the tank due to the high operating temperatures of the SOFC.

MacDonald et al. [12] studied the behavior of a thermally coupled metal hydride tank containing $Ti_{0.98}Zr_{0.02}V_{0.43}Fe_{0.09}Cr_{0.05}Mn_{1.5}$ alloy with a Proton Exchange Membrane fuel cell (PEMFC), by using three different geometries for the tank. They exported that the annular tank could provide hydrogen gas at the required flow rate and pressure stimulated by the fuel cell. Jiang et al. [13] studied the behavior of a thermally coupled hydrogen storage tank and a PEMFC, where the effect of transferring waste thermal energy from the fuel cell to the hydride was examined. They pointed the importance of providing energy to the metal hydride bed in order to facilitate the desorption process. Delhomme et al. [14] presented an experimental setup designed to test the thermal integration of a $MgH_2$ tank with an operating SOFC. It was extracted that using the heat exhaust from the SOFC to provide heat to the hydride bed is feasible only as long as the gas is found at a sufficiency high temperature (above 350-400 $^0C$) to the tank.

Among all the metals used for the formation of hydrides, Magnesium has the highest energy density and the storage capacity that presents is 7.6 wt%. The main drawback of using $MgH_2$ as a hydrogen storage vessel is the requirement of medium grade heat (350 $^0C$) for the desorption process [15-17]. In order to maintain or even optimize the desorption of hydrogen from a $MgH_2$ tank, the coupling and thermal integration with the SOFC seems a very promising option. Heat is released from the SOFC stack at high temperature (800-850 $^0C$) [18, 19] and can be used to enhance the release of hydrogen within the $MgH_2$ tank

The current study aims to improve the understanding of the hydrogen desorption process within $MgH_2$ storage system by taking account and simulating four different scenarios for each hydride. The first scenario deals with the simulation of the desorption process of the hydrides without taking into account the coupling with the SOFC, where the energy for the endothermic desorption is given by an external heater. The second scenario involves, for the very first time, according to the current knowledge of the authors, the effect of the irradiation heat transfer from the high temperature operating SOFC to the desorbing metal hydride bed. For the third scenario, the exhaust air from the operating SOFC is driven to the metal hydride with the use of pipes and finally, a combination of the external heater and exhaust gas heat transfer is studied in order to decide which of the scenarios have the greater efficiency at the hydrogen desorption process. The above scenarios expose the advantages of providing waste heat energy from the SOFC stack to the metal hydride bed in order to facilitate the removal of hydrogen. For the current study, the geometry of the SOFC and the metal hydride tank was implemented in a finite element program called COMSOL Multiphysics.

2. Mathematical Modeling of the Hydrogen Desorption Process.

*2.1 Geometry of the Metal Hydride Tank*

The metal hydride bed used in the current study, consists of a cylindrical container of radius R=0.05m and height H=0.5m which contains the metal powder that chemically absorbs the hydrogen under pressure. For the simulation studies, four different geometries are considered, based on the cylindrical container. The first geometry consists of the cylindrical container with a small cylinder at the basis, which is the hydrogen supply canister, and the heat source is an external heater. The second geometry is also a cylindrical container with the same dimensions while in this case the heat source is the heat flux from the radiation from an operating SOFC. The third geometry is a cylindrical container thermally connected via pipes to the operating SOFC where the exhaust heat from the SOFC is passing the metal hydride through a co-axial heat exchanger to the cylindrical container and the final geometry consists of a thermally coupled SOFC and the metal hydride tank with pipes, while the heat source is a combination of both external heater and the exhaust heat from the SOFC. Fig. 1 illustrates the four different geometries under operating conditions, where the hydrogen desorption is taking place.

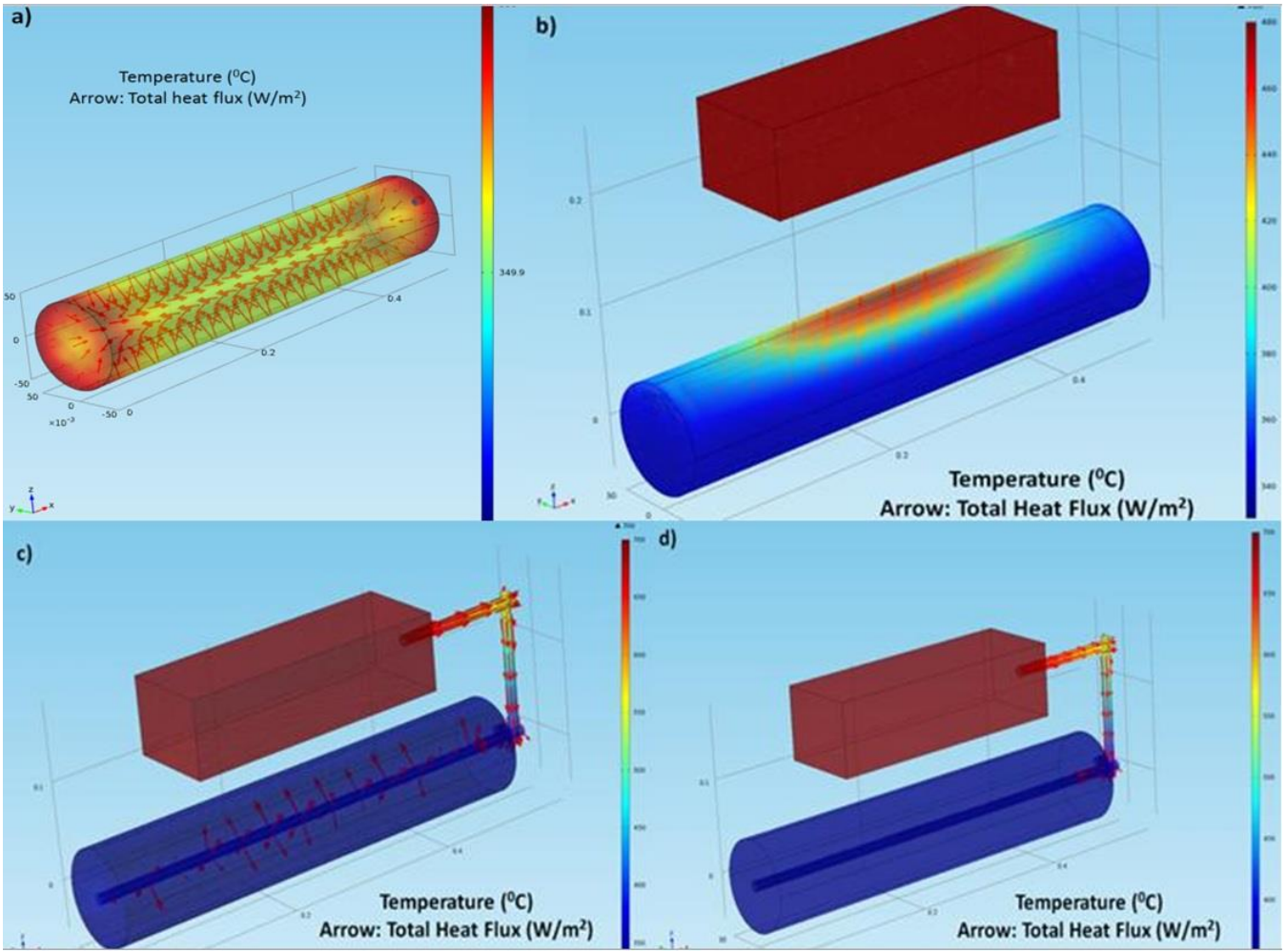

*Fig. 1. Geometries of the simulation study. Fig 1a. represents the hydrogen desorption by using the external heater as the heat source for desorption. Fig 1b. illustrates the use of radiation heat from the SOFC as heat source for desorption. Fig 1c. shows the geometry with the exhaust heat from SOFC as the heat source and Fig. 1d shows the geometry where both external heater and the exhaust from the SOFC are used as the heat source for the desorption.*

*2.2 Problem Formulation*

The governing equations, consist of energy, mass and momentum conservation which described by partial differential equations, and some other equations that describe the kinetics of absorption and desorption [20, 21]. The model describes also the diffusion through the hydride bed, by using the Darcy's law and taking into consideration the kinetics of absorption as a function of the difference between the local and equilibrium temperature.

In order to study the hydrogen desorption process in a metal hydride bed, simulations are performed using COMSOL Multiphysics 4.2, as it is specialized in solving the coupled heat and mass transfer problems in porous media. Further, this software takes into account the gas transport (diffusion and flow) pressure gradients inside the hydride beds and the reaction kinetic inside the alloys. In the current simulation study Mg powder is considered inside the metal hydride tank. The operating conditions and physical properties of magnesium hydride and hydrogen at 600 K, are presented in Table 1.

*Table 1. Operating conditions and physical parameters of magnesium hydride and hydrogen*

| Parameter | Value |
|---|---|
| **Desorption Enthalpy $\Delta H$** | -75500 (J/mol) |
| **Desorption Entropy $\Delta S$** | -135.6 (J/K/mol) |
| **Porosity $\varepsilon$** | 0,654 |
| **Mg Heat Capacity $Cp$** | 1545 (J/kg/K) |
| **Mg Thermal Conductivity $\lambda$** | 0,48 (W/m/K) |
| **Mg Density $\rho_{Mg}$** | 1800 (kg/m$^3$) |
| **Initial Temperature $T_0$** | 320 ($^0$C) |
| **Initial Pressure $P_0$** | 0.8 (bar) |
| **External Heater Temperature $T_{inf}$** | 350 ($^0$C) |
| **SOFC Operating Temperature $T_{op}$** | 700 ($^0$C) |
| **Hydrogen Dynamic Viscosity $\mu$** | 8.6 10$^{-6}$ (Pa s) [30] |
| **Hydrogen Molecular Mass $M_{gas}$** | 0.002 (kg/mol) |
| **Hydrogen Thermal Conductivity $\lambda_{H2}$** | 0.18 (W/m/K) |

The modules added in the current model are the heat transfer in porous media module which includes the energy conservation equation, the Darcy's law module including the momentum conservation equation and the transport in diluted species module containing the mass conservation equation. In order to simplify the problem, some assumptions made such as:

- The medium (gas and metal) initially are in local thermal equilibrium
- The solid phase is isotropic and has uniform porosity
- In order to be able to calculate for every temperature the hydrogen density, the hydrogen is treated as an ideal gas
- The thermophysical properties of the hydride bed are independent of the bed's temperature and the hydrogen supply temperature
- The effect of hydrogen conservation on the variation of the equilibrium pressure is negligible.

*2.2.1 Energy Conservation Equation*

Assuming thermal equilibrium between the hydride powder and hydrogen, a single energy equation is solved instead of separate equations foe both solid and gaseous phases.

$$(\rho \cdot Cp)_e \cdot \frac{\partial T}{\partial t} + (\rho_g \cdot Cp_g) \cdot \bar{v}_g \cdot \nabla T = \nabla \cdot (k_e \cdot \nabla T) + m \cdot (\Delta H - T \cdot (Cp_g - Cp_s)) \quad (1)$$

Considering only parallel heat conduction in solid and gas phases, there are the following expressions for specific heat and thermal conductivity respectively:

$$(\rho \cdot Cp)_e = (\varepsilon \cdot \rho_g \cdot Cp_g) + ((1-\varepsilon) \cdot \rho_s \cdot Cp_s) \quad (2)$$

$$k_e = \varepsilon \cdot k_g + (1-\varepsilon) \cdot k_s \quad (3)$$

Both the equations (2) and (3) are expressed as porosity – weighted functions of the hydrogen – gas and the solid – metal phases.

## 2.2.2 Hydride Mass Balance

For the solid, a mass conservation equation is considered.

$$(1-\varepsilon)\frac{\partial(\rho_s)}{\partial t} = -m \qquad (4)$$

## 2.2.3 Hydrogen mass balance

The mass conservation for the gas is considered as:

$$\varepsilon \cdot \frac{\partial(\rho_g)}{\partial t} + div(\rho_g \cdot \vec{v}_g) = -m \qquad (5)$$

## 2.2.4 Momentum equation

The gas velocity can be expressed using Darcy's law. By neglecting the gravitational effect, the equation is the above:

$$\vec{v}_g = -\frac{K}{\mu_g} \cdot grad(\vec{P}_g) \qquad (6)$$

Where K is the permeability of the solid and $\mu_g$ is the dynamic viscosity of gas. The solid permeability is given by the Kozeny – Carman's equation:

$$K = \frac{dp^2 \cdot \varepsilon^3}{180 \cdot (1-\varepsilon^2)} \qquad (7)$$

Assuming that the hydrogen is an ideal gas, from the perfect gas law ($\rho_g = (P_g M_g)/(RT)$) and considering Darcy's law, the mass conservation equation of hydrogen becomes:

$$\frac{\varepsilon \cdot M_g}{R \cdot T} \cdot \frac{\partial P_g}{\partial t} + \frac{\varepsilon \cdot M_g \cdot P_g}{R \cdot T} \cdot \frac{\partial}{\partial t} \cdot \frac{1}{T} - \frac{K}{v_g \cdot r} \cdot \frac{\partial}{\partial r} \cdot \frac{r \cdot \partial P_g}{\partial r} - \frac{K}{v_g} \cdot \frac{\partial^2 P_g}{\partial z^2} = -m \qquad (8)$$

## 2.2.5 Desorption Kinetic Expression

For the desorption process, the following expression is used in order to describe the kinetics.

$$m = C_d \cdot \exp[-\frac{E_d}{R_g \cdot T}] \cdot (\frac{P_{eq} - p_g}{P_{eq}}) \cdot (\rho_s - \rho_o) \qquad (9)$$

Where m is the source term and used in equations (1), (4), (5), (8). $C_d$ is pro-exponential constant for desorption, $E_d$ is the activation energy for desorption, $\rho_s$ is the solid density, and $\rho_o$ is the initial metal hydride density.

## 2.2.6 Equilibrium Pressure

The equilibrium pressure for the hydrogen, which is the most important parameter which defines if the reaction is going to take place or not, is given by van't Hoff law:

$$\ln P_{eq} = \frac{\Delta H}{R_g \cdot T} - \frac{\Delta S}{R_g} \qquad (10)$$

*2.3 Model Validation*

When simulating a problem, verification of the proposed model and the simulation results is required to ensure that the model does what is intended to do. Validation of the current model has been performed by comparing the extracted simulation results to experimental results already been published. Figure 2 shows the comparison between the experimental results by Dornheim et al. [22] and the simulating results according to the proposed model for three different cases. For hydrogen desorption from a pure-Mg hydride tank, for nanocrystalline Mg hydride tank and for a $Nb_2O_5$ catalyzed Mg hydride tank. As extracted from the results the desorption curves presents similar behavior for all the cases, including very slow kinetics for the pure Mg and much faster for the other two cases.

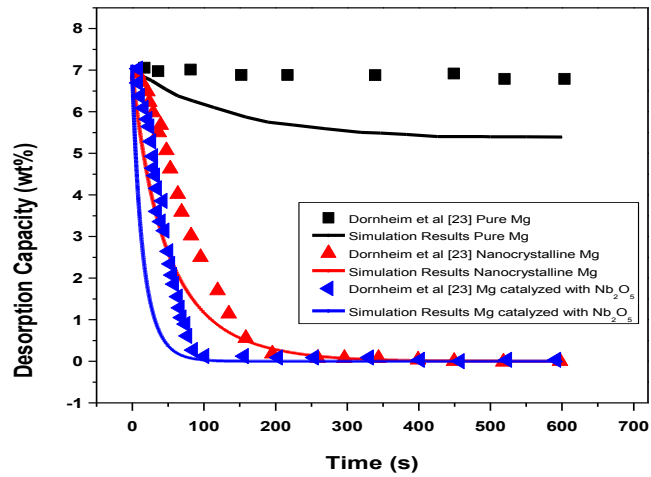

*Fig. 2. Validation of the proposed model with the experimental results by Dornheim et al [22].*

3. Radiation Heat Transfer Modeling

Generally, SOFCs operates at high temperatures between 600-1000 $^0C$ [23] and due to the high operating temperatures, radiation heat transfer must be given special consideration in thermal modeling efforts. Damn et al. [24] proposed a model of radiative heat transfer within the SOFC, assuming that the outer surface of the SOFC stack as an isothermal surface, in order to maintain the temperature inside the stack. In the current work the radiation heat transfer from the outer surface of the stack to the surface of the metal hydride tank is simulated, in order to use the heat origin from the operating SOFC as the heat source to maintain the desorption process.

The governing equations for radiative heat transfer are integral-differential equations, and they are not linear, as the emissive power features a fourth-power dependence on the temperature. Many researchers used different methods to model the radiation heat transfer [25-28]. In order to propose a single mathematical model, the SOFC's outer surface and the hydrides outer surface are assumed as opaque, diffuse and grey surfaces. The radiosity, represents the rate at which radiation energy leaves a unit of a surface in all directions. For an opaque surface, the radiosity is expressed by:

$$J = (1 - \varepsilon_m)G + \varepsilon_m \sigma T^4 \qquad (11)$$

Where J is the radiosity (W/m$^2$), $\varepsilon_m$ is the surface emissivity, G is the irradiation (W/m$^2$) and σ is the Stefan-Boltzmann constant (5.67 10$^{-8}$ W/m$^2$K$^4$). The radiation heat flux incident on a surface from all directions is called irradiation G and is expressed as [29]:

$$G = G_m + F_{amb}\sigma T_{amb} \quad (12)$$

Where $G_m$ is the mutual irradiation coming from the other surfaces in the model and is expressed as:

$$G_m = \sum_{j=1}^{N} \frac{(-\vec{n}\vec{r})(n\vec{r})}{n|r|^2} \quad (13)$$

The above equations, Eq. (11), Eq. (12) and Eq. (13) are used in order to calculate the radiation flux from the operating SOFC to the metal hydride tank.

4. Results and Discussion

For the SOFC operation, it is assumed that the SOFC is servicing an intermittent load cycling with a 30 min period. The on-cycle then lasts for 15 min and this is the time range where the thermal coupling effect between the operating SOFC and the metal hydride tank will be examined. Further, four different scenarios about the thermal coupling will be simulated and discussed. The first scenario is a "reference" scenario, where the heat for the desorption process within the tank is offered to the system by an external heater. The second scenario uses the heat from the irradiation heat flux from the operating SOFC to the metal hydride tank. The third scenario involves the exhaust heat from the operating SOFC to the metal hydride bed and the fourth scenario is a combination between the exhaust heat from the operating SOFC to the metal hydride bed and the external heater in order the hydride tank to release the hydrogen. Further, for the second scenario, a separate study will be performed in order to clarify the role of the radiative heat transfer from the SOFC to the hydride tank. The temperature and hydrogen concentration profiles across the z-axis of the metal hydride tank will be examined and discussed and finally five different "thermocouples" named as TC1, TC2, TC3, TC4, TC5 were assumed inside the hydride tank in order to be able to measure the temperature, pressure and concentration inside the tank. The position of these "thermocouples" is shown if Fig. 3.

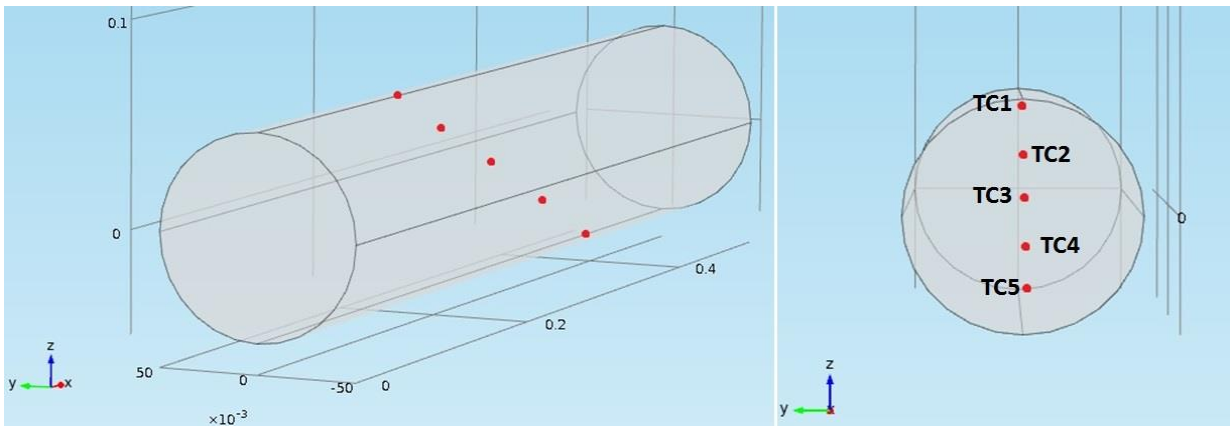

*Fig. 3. The position of the five "thermocouples" inside the metal hydride tank*

*4.1 Hydrogen desorption with the heat from the external heater.*

According to Table 1, the initial temperature of the hydride tank is 320 $^0$C and the external heater is at 350 $^0$C. The simulation runs were performed for 15 min (900 s), which is the time where the SOFC last for the on-cycle. Fig. 4 presents the temperature profile of the metal hydride tank across the z-axis of the tank as shown in Fig. 3. The temperature deviation is not uniform within the hydride tank. The heat is provided at the external surfaces of the tank from the external heater and due to the low thermal conductivity of the metal hydride, the temperature profile drops until the heat reaches the zone where the reaction is taking place. It is also notable that the lower temperature is at the center of the hydride tank indicating that the desorption process has a preferred direction from the outer of the tank to the middle of the tank.

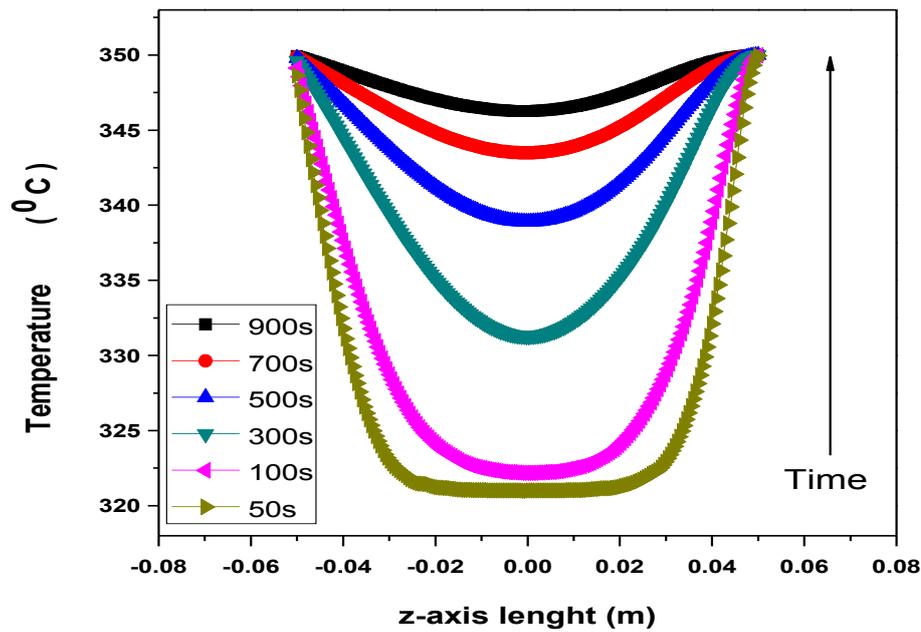

*Fig. 4. Temperature profiles across the z-axis of the hydride tank for different times for the hydrogen desorption with the heat provided by external heater.*

Fig. 5 illustrates the concentration of the desorbed hydrogen within the metal hydride tank along the z-axis. It is notable that the desorption process is symmetrical with the symmetry axis to be at the center of the hydride tank and confirms the previous assumption that the desorption process has a preferred direction within the tank.

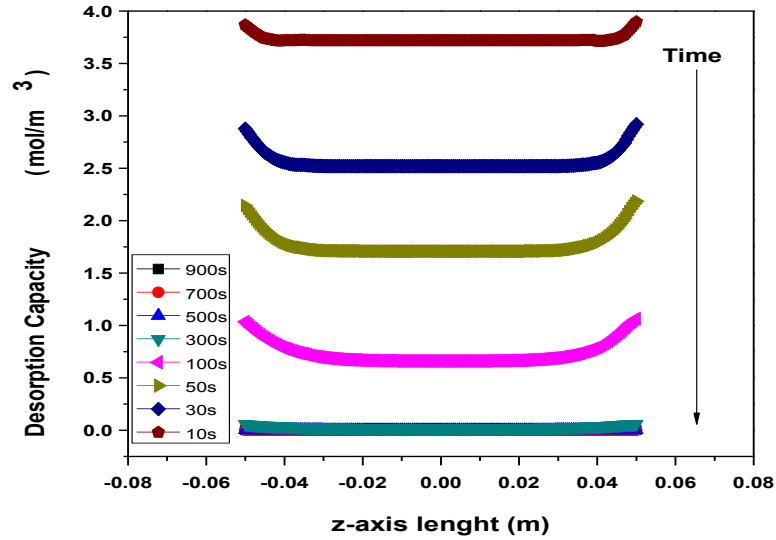

*Fig. 5. Hydrogen concentration profiles across the z-axis of the hydride tank for different times for the hydrogen desorption with the heat provided by external heater.*

4.2 Hydrogen desorption with the heat from the radiation flux from the operating SOFC.

Due to the high operating temperature of SOFC, the radiative heat transfer from the SOFC to the hydride is a crucial parameter and can be used in order to enhance the desorption process. According to [25], heat losses from the edges of the SOFC have the potential to induce damaging thermal gradients within the cells. It was also insisted that little has been reported about the interaction between the SOFC stack and other surfaces outside the stack and the role of radiation heat transfer in minimizing heat losses from the stack. In the current study, it is assumed that the SOFC emits radiative heat transfer without resulting damaging thermal gradients within the cells.

*4.2.1 Identifying appropriate conditions for radiative heat transfer from the operation SOFC to the metal hydride tank*

It is well known that the melting point of pure Mg is 650 $^0$C. In the current simulation study it is assumed that the SOFC operates at 700 $^0$C. It is therefore of major concern the avoidance of heating the material above 350-400 $^0$C, which is a temperature range capable to maintain the desorption process of hydrogen within the metal hydride tank. The first step when examine the effect of radiative heat transfer from the operating SOFC to the hydride tank is to discover the conditions at which the radiation heat transfer doesn't damage the material (distance between SOFC and metal hydride, operating temperature, time of radiation flux). Fig. 6 presents the effect of the radiative heat transfer between the SOFC and the metal hydride tank after 900s of radiation where the distance between them is 0.13m and the operating temperature of the SOFC is 700 $^0$C.

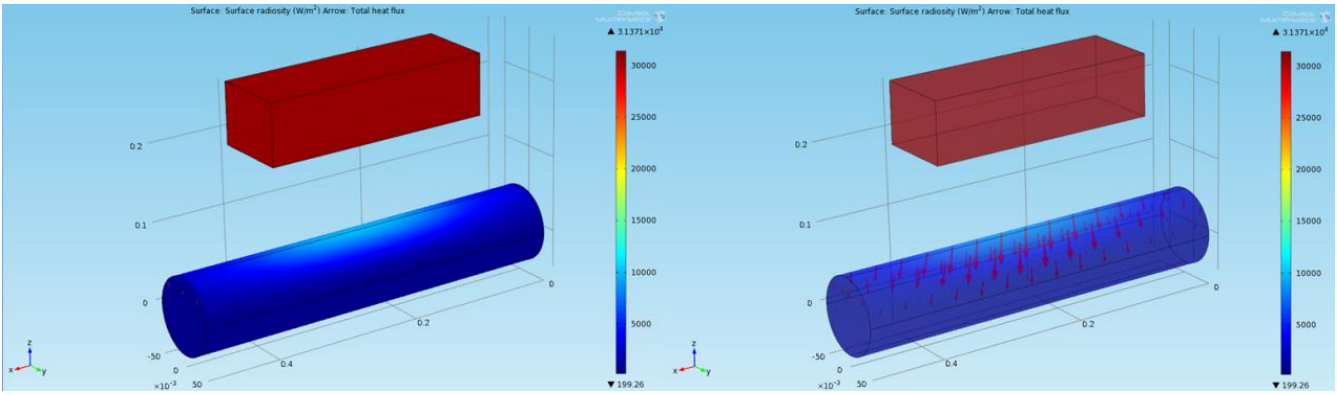

*Fig. 6. Surface radiosity after 900s of operation of the SOFC at 700 $^0$C. The arrows represent the direction of the total heat flux.*

The arrows inside the metal hydride tank represent the direction of the total heat flux within the tank and it is obvious that the heat flux goes from the outer surface exposed to the radiation of the SOFC to the center of the tank. Further, Fig. 7a shows the total energy flux profiles for the five "thermocouples" within the tank and Fig. 7b shows the temperature profile for the "thermocouples". According to these results, it seems that the maximum temperature within the tank is 430 $^0$C for the TC1 while for the others the temperature varies from 332-375 $^0$C which is a temperature range capable to maintain the desorption process. These conditions are potential for using the radiation heat transfer for the heat source for the release of hydrogen within the tank and will be used for the simulation of the dehydrogenation process.

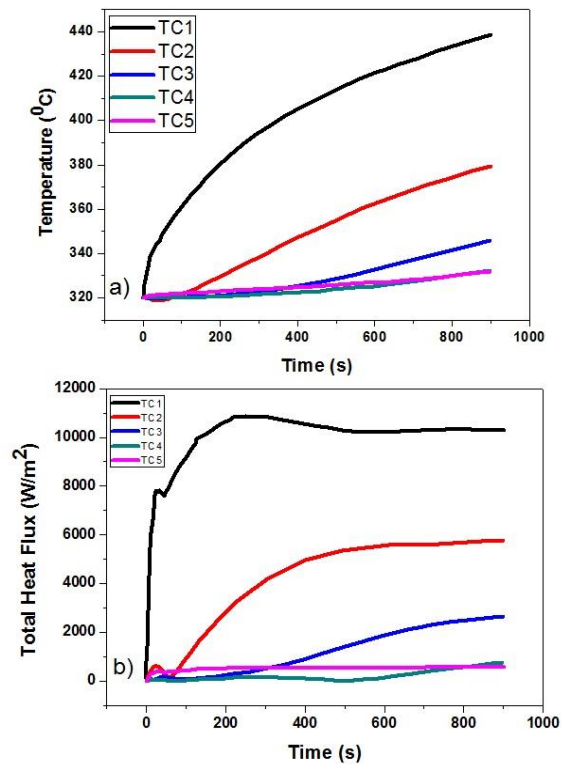

*Fig. 7. Temperature and Total Heat Flux profiles for all the thermocouples inside the hydride tank indicating the effect of radiative heat transfer inside the hydride*

*4.2.2 Hydrogen desorption with the radiation heat transfer as heat source*

The desorption process within the tank is studying, and as the heat source the radiation heat transfer from the operating at 700 $^0$C SOFC to the hydride tank is used. Fig. 8 illustrates the temperature profile across the z-axis of the cylindrical hydride tank at different desorption times.

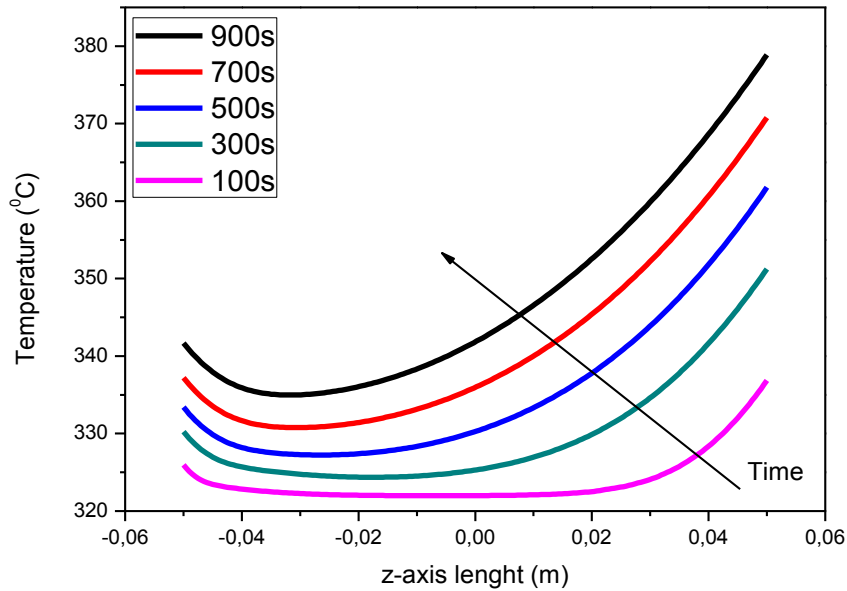

*Fig. 8. Temperature profiles across the z-axis of the hydride tank for different times for the hydrogen desorption with the heat provided by the radiation heat transfer.*

As extracted from the results, the temperature profile has a non-uniform distribution where the lower value of temperature is located in the space within the metal hydride near the surface which has the lowest radiation exposure. The highest temperature is located near the surface which exposed directly to the radiation from the SOFC. Further, from Fig. 9, which shows the concentration of desorbed hydrogen within the tank, the highest concentration of the desorbed hydrogen is found near the surface where the radiation flows directly. Inside the tank, the hydrogen concentration appears to have a uniform distribution with almost the same value across the axis, with a small rise near the surface with the lowest radiation exposure. The results showed that in this case the desorption process has also a preferred direction from the outer surface, directly exposed to the radiation to the center and from all the other surfaces not directly exposed to the radiation to the center of the tank. In the case of the heat flux from the directly exposed surface the rate of the heat transfer is higher than from the other surfaces.

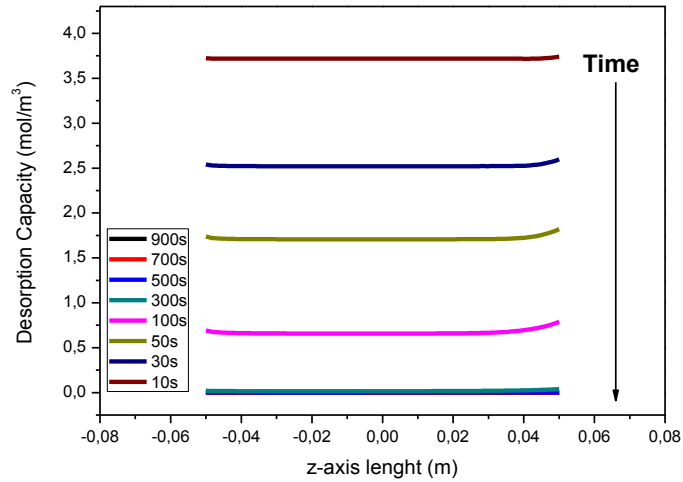

*Fig. 9. Hydrogen concentration profiles across the z-axis of the hydride tank for different times for the hydrogen desorption with the heat provided by radiation heat transfer.*

4.3 Hydrogen desorption with the heat from the exhaust gas from the operating SOFC.

Under experimental conditions, usually the SOFC cathode inlet air is supplied from a compressed air supply line and the cathode stream is pre-heated before entering the cathode distribution manifold within the stack [14]. The cathode exhaust is vented with the SOFC operation only, and during the thermal coupling with the hydride tank, the cathode exhaust is mixed with fresh air at 20 $^0$C, instead to control the temperature of the stream entering the hydride vessel at 350 $^0$C and thus, avoiding a partial sintering of the magnesium in the tank. In this case, is assumed that the temperature of the exhaust when entering the metal hydride tank is 350 $^0$C and remains constant for the 15 min of the on-cycle of the SOFC. The exhaust air from the SOFC is driven with pipes to the metal hydride tank as shown in Fig. 1c. The pipe inside the metal hydride tank is placed at the center of the cylindrical tank and also the pipes have cylindrical shape. Fig. 10 illustrates the temperature profile within the metal hydride tank across z-axis for different desorption times. It is extracted that the temperature profile in this case has a symmetrical shape with center of symmetry the center of the tank where the pipe with the exhaust air is placed. As the time is passing, the temperature within the tank is increased, but cannot reach the maximum value. This is probably due to the poor thermal conductivity. The maximum temperatures are found near the center of the tank where the heat source has been placed and the lower temperature values are found near the outer surfaces of the tank. Fig. 11 presents the desorbed hydrogen concentration within the tank at various desorption times. The concentration also seems to be symmetrical within the tank and has almost the same value for all the domains inside the tank. The concentration profile shows that during the presence of a heating medium in the middle of the tank, the concentration of the desorbed hydrogen within the tank is homogeneous and have almost the same concentration everywhere inside the tank.

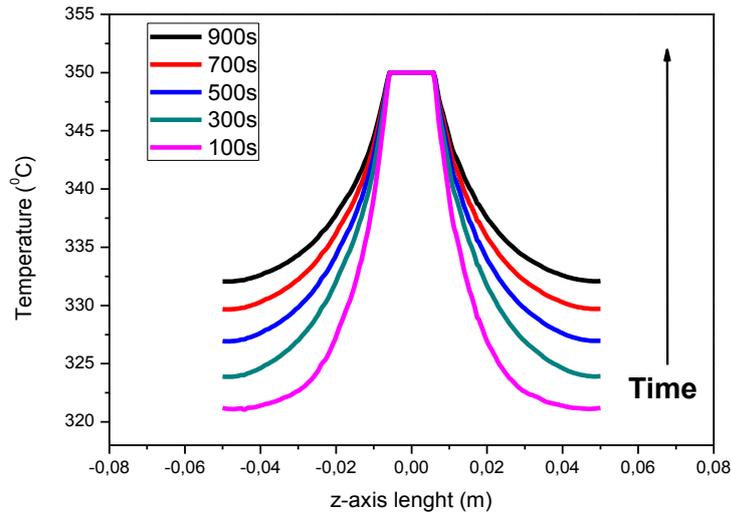

*Fig. 10. Temperature profiles across the z-axis of the hydride tank for different times for the hydrogen desorption with the heat provided by the exhaust from the SOFC.*

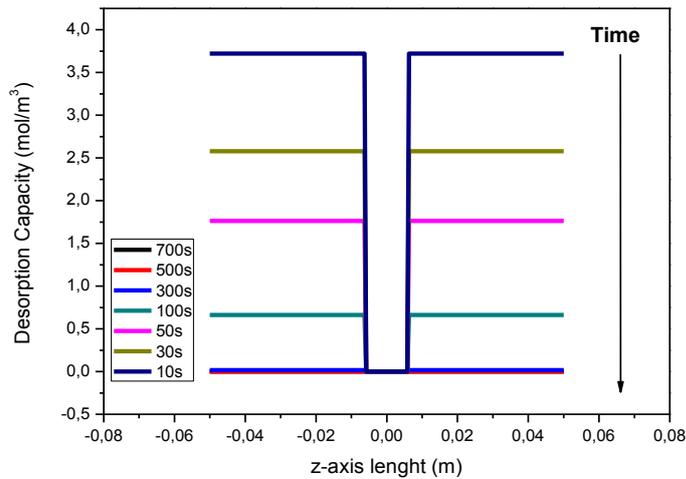

*Fig. 11. Hydrogen concentration profiles across the z-axis of the hydride tank for different times for the hydrogen desorption with the heat provided by the exhaust from the SOFC.*

4.4 Hydrogen desorption with the combination heat from the external heater and the exhaust heat from the operating SOFC.

According to the last scenario of the current simulation study, the heat for the release of hydrogen from the chemical bonds with the Mg atoms is provided to the metal hydride tank from both an external heater (350 $^0$C) and the exhaust air from SOFC (350 $^0$C) as described at chapter 4.3. The external heater provides heat to the outer surfaces of the tank while the exhaust air provides heat internally. Fig. 12 shows the temperature profile across the z-axis of the cylindrical hydride tank for different times. The temperature distribution in this case presents symmetry with two different symmetry centers, the first one in between the center of the hydride tank and the left outer surface and the second in between the center of the hydride and the right outer surface. As time is passing, the temperature inside the hydride tank increases while the symmetry remains. At the end of the 900s it seems that the temperature within the tank is everywhere almost the same.

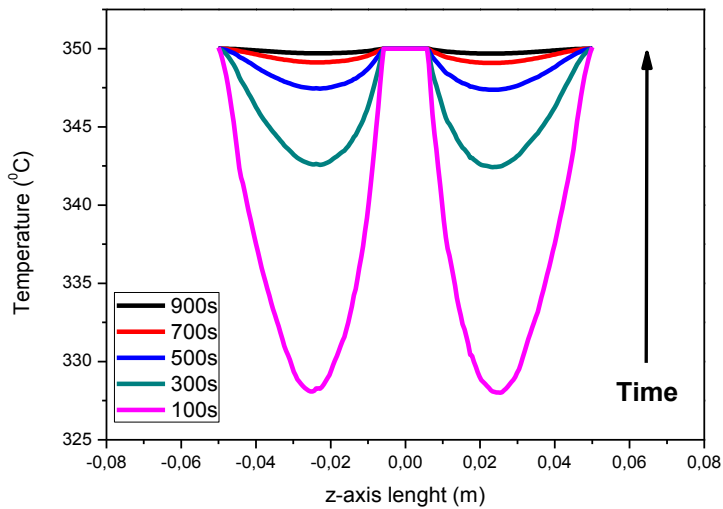

*Fig. 12. Temperature profiles across the z-axis of the hydride tank for different times for the hydrogen desorption with the heat provided by the combination of the exhaust from the SOFC and the external heater.*

Fig. 13 also shows that the concentration of the desorbed hydrogen within the metal hydride tank is homogeneous with a symmetry center in the middle of the tank as discussed in the case of the heating only with the exhaust heat gas from the SOFC.

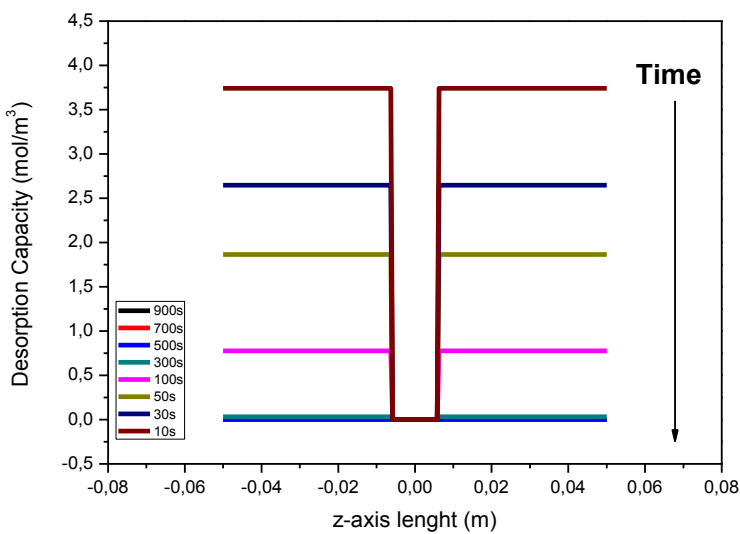

*Fig. 13. Hydrogen concentration profiles across the z-axis of the hydride tank for different times for the hydrogen desorption with the heat provided by the combination of the exhaust from the SOFC and the external heater.*

4.5 Comparison of temperature behavior within the metal hydride tank for all the simulating scenarios

Fig. 14 illustrates a comparison of the temperature deviation within the metal hydride tank for the five thermocouples (TCs) as defined earlier. The comparison is performed for all the coupling scenarios simulated in the current study and the results are taken after 900s of desorption process, at the end of the on-cycle of the SOFC.

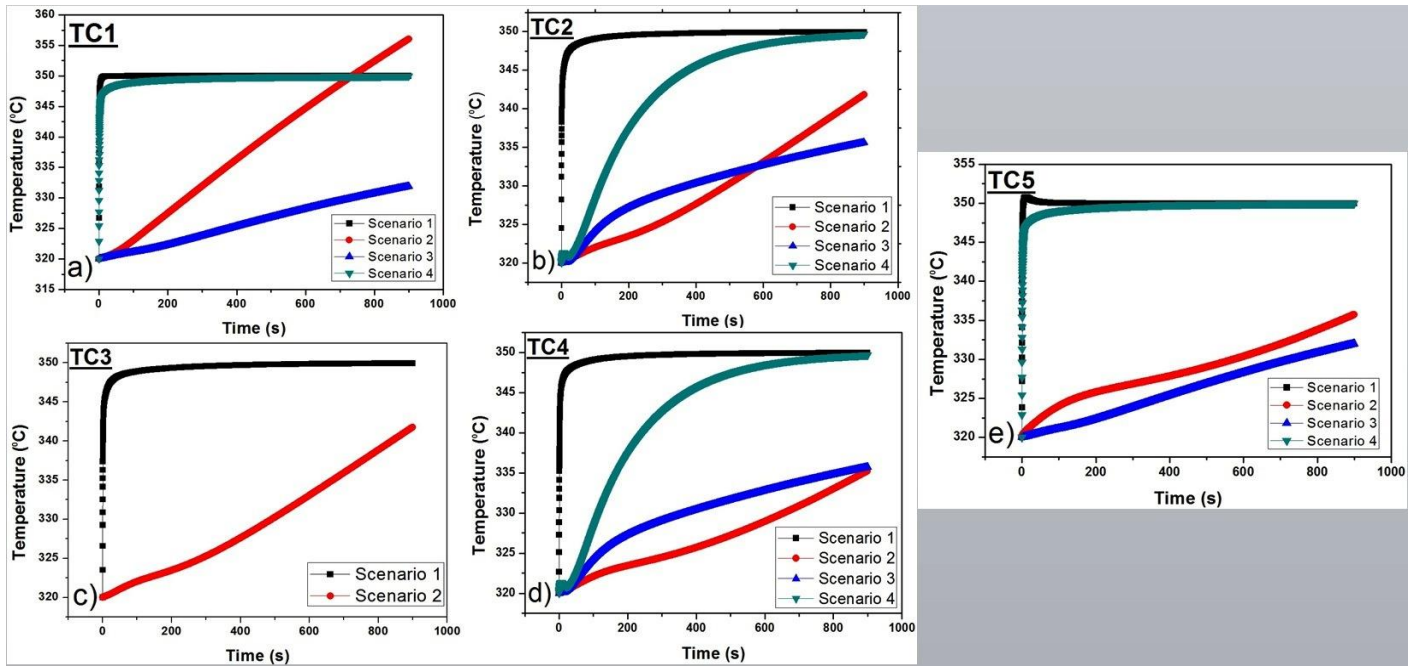

*Fig. 14. Comparison of temperature deviation within the metal hydride tank. Fig. 14a shows the temperature profile for the TC1 for all the simulating scenarios, while Fig. 14b, Fig. 14c, Fig. 14d and Fig. 14e shows the temperature scenario for the TC2, TC3, TC4 and TC5 respectively.*

Temperature is increased faster at the first scenario where the external heater is consider as the heat source, indicating that after the 900s the temperature within the tank has almost reached the external heater temperature. The increase of the temperature is more pronounced at the outer surfaces of the tank and more specifically at TC1 and TC5. For the radiation heat used as a heat source (second scenario) the temperature rises relatively slow, which is explained by the mechanism, which is responsible for the temperature increase. It should be noted that the temperature is higher and the increase rate also higher to the TCs which are closer to the surface, which is directly exposed to the radiation flux from the SOFC. For the scenarios which used the exhaust heat from the SOFC as the heat source (scenario 3) and the combination between the exhaust and the external heater (scenario 4), the temperature increases relatively slow (scenario 3) and fast (scenario 4). For the third scenario the temperature is higher to the TC2 and TC4 which are closer to the heating pipe at the center of the metal hydride tank.

5. Conclusions

Thermal coupling between a Mg metal hydride tank and an operating SOFC at 700 $^0$C was simulated. The main objective of the current study was to investigate the behavior of the desorption process of the hydride under different heat sources in order to find the best way to maintain the temperature for this endothermic reaction. This objective was met by formulating a dynamic dehydriding model which was validated by comparing the results extracted from the model with some experimental data already been published. Four scenarios were proposed in order to compare the ability of the dehydriding process, by using as heat sources an external heater, the heat from the radiation for the SOFC, the exhaust air from the SOFC and a combination of an external heater with the exhaust air from the SOFC. The results revealed that the thermal coupling of the SOFC with the hydride tank is capable to maintaining the temperature within the hydride tank in almost homogeneous levels everywhere inside the tank, especially in the case of the combination of both external heater with the exhaust heat from the SOFC. In the case of the radiative heat transfer from the SOFC, the temperature profile presents an almost parabolic distribution with the higher temperature to be near the surfaces which are directly exposed to the radiation.

**Nomenclature**

| | | | | |
|---|---|---|---|---|
| $C_p$ | specific heat capacity, J/kg/K | | | *Greek letters* |
| $T$ | temperature, K | | $\varepsilon$ | porosity |
| $\bar{v}$ | gas velocity, m/s | | $\varepsilon_m$ | surface emissivity |
| $k$ | kinetic coefficient, $s^{-1}$ | | $\sigma$ | Stefan-Maxwell constant, (5.67 $10^{-8}$) $Wm^{-2}K^{-4}$ |
| $m$ | kinetic expression for desorption | | $\rho$ | density, kg/$m^3$ |
| $K$ | permeability, $m^2$ | | $\mu$ | dynamic viscosity, Pas |
| $P$ | hydrogen pressure, Pa | | | |
| $R$ | universal gas constant, $JK^{-1}mol^{-1}$ | | | *subscripts* |
| $C$ | desorption constant, $s^{-1}$ | | g | gas, hydrogen |
| $E$ | desorption activation energy, $Jmol^{-1}$ | | s | solid, metal |
| $P_{eq}$ | equilibrium pressure, Pa | | d | desorption |
| $\Delta H$ | molar enthalpy for desorption, $Jmol^{-1}$ | | eq | equilibrium |
| $\Delta S$ | molar entropy for desorption, $Jmol^{-1}K^{-1}$ | | amb | ambient |
| $M$ | hydrogen molar mass, $kgmol^{-1}$ | | e | efficient |
| $J$ | radiosity, $Wm^{-2}$ | | | |
| $G$ | irradiation flux, $Wm^{-2}$ | | | *operators* |
| $F$ | view factor | | $\nabla$ | gradient or nabla |

*n*